# Cross-Modality Learning for Predicting IHC Biomarkers from H&E-Stained Whole-Slide Images

Amit Das[1], Naofumi Tomita, MS[2], Kyle J. Syme, MD[3], Weijie Ma, MD[3], Paige O'Connor, MD[3], Kristin N. Corbett[3], Bing Ren, MD, PhD[3], Xiaoying Liu, MD[3†], Saeed Hassanpour, PhD[1,2,4†*]

[1]Department of Computer Science, Dartmouth College, Hanover, NH 03755, USA
[2]Department of Biomedical Data Science, Geisel School of Medicine at Dartmouth, Hanover, NH 03755, USA
[3]Department of Pathology and Laboratory Medicine, Dartmouth-Hitchcock Medical Center, Lebanon, NH 03756, USA
[4]Department of Epidemiology, Geisel School of Medicine at Dartmouth, Hanover, NH 03755, USA
† Xiaoying Liu and Saeed Hassanpour contributed equally as senior authors.
*Corresponding Author: Saeed Hassanpour, PhD
Postal address: One Medical Center Drive, HB 7261, Lebanon, NH 03756, USA
Email: Saeed.Hassanpour@dartmouth.edu

**Conflicts of Interest**
The authors have no financial, professional, or personal conflicts of interest.

**Funding**: This research was supported in part by grants from the US National Library of Medicine, R01LM012837 (S.H.) and R01LM013833 (S.H.), and the US National Cancer Institute, R01CA249758 (S.H.).

# ABSTRACT

Hematoxylin and Eosin (H&E) staining is a cornerstone of pathological analysis, offering reliable visualization of cellular morphology and tissue architecture for cancer diagnosis, subtyping, and grading. Immunohistochemistry (IHC) staining provides molecular insights by detecting specific proteins within tissues, enhancing diagnostic accuracy, and improving treatment planning. However, IHC staining is costly, time-consuming, and resource-intensive, requiring specialized expertise. To address these limitations, this study proposes *HistoStainAlign*, a novel deep learning framework that predicts IHC staining patterns directly from H&E whole-slide images (WSIs) by learning joint representations of morphological and molecular features. The framework integrates paired H&E and IHC embeddings through a contrastive training strategy, capturing complementary features across staining modalities without patch-level annotations or tissue registration. The model was evaluated on gastrointestinal and lung tissue WSIs with three commonly used IHC stains: P53, PD-L1, and Ki-67. HistoStainAlign achieved weighted F1 scores of 0.735 [95% Confidence Interval (CI): 0.670–0.799], 0.830 [95% CI: 0.772–0.886], and 0.723 [95% CI: 0.607–0.836], respectively for these three IHC stains. Embedding analyses demonstrated the robustness of the contrastive alignment in capturing meaningful cross-stain relationships. Comparisons with a baseline model further highlight the advantage of incorporating contrastive learning for improved stain pattern prediction. This study demonstrates the potential of computational approaches to serve as a pre-screening tool, helping prioritize cases for IHC staining and improving workflow efficiency.

**INTRODUCTION**

Microscopic examination of stained tissue slides is the gold standard for diagnosing a wide range of diseases, including cancer. Hematoxylin and Eosin (H&E) staining is the most widely used technique in pathology due to its ability to reveal cellular morphology and tissue architecture. This staining procedure is critical in cancer diagnosis, subtyping, and grading, owing to its widespread availability, reliability, and cost-effectiveness[1].

In pathology, immunohistochemistry (IHC) staining is an auxiliary laboratory technique increasingly used to identify and visualize specific proteins within formalin-fixed paraffin-embedded (FFPE) tissue samples. This method employs antibodies that bind to target proteins, which are then highlighted using chromogens of varying colors. By providing molecular insights into the presence and spatial distribution of specific molecules, IHC staining significantly enhances diagnostic accuracy and disease classification, particularly in cancers, compared to solely morphological information derived from standard H&E staining alone[2].

Several IHC stains, including P53, Ki67, and PD-L1, are routinely used for cancer diagnosis, prognostication, and predictive testing to assess responses to targeted therapies such as immunotherapy. P53, a tumor suppressor protein that regulates cell division and apoptosis, is one of the most commonly mutated proteins in malignant tumors. In its mutated form, P53 becomes more stable and overexpressed in cancer cells; its expression can be detected through P53 IHC staining[3]. Programmed death ligand-1 (PD-L1), a protein that binds to PD-1 on activated immune cells, suppresses immune responses against cancer cells. Immune checkpoint inhibitors can block PD-L1, reactivating immune cells to target cancer. PD-L1 expression, detectable via IHC, is a predictive biomarker for immunotherapy response in cancers[4]. Similarly, Ki-67, a nuclear protein associated with cell proliferation, serves as a key marker for tumor grading, prognostication, and targeted cancer therapy[5].

Despite their utility, IHC stains have several limitations, including high costs, time and labor-intensive protocols, tissue exhaustion, and the need for specialized expertise and equipment for interpretation. Inter-observer variability further complicates their applications, potentially affecting diagnostic consistency. As a result, there is growing interest in leveraging advanced computational methods, such as deep learning (DL), to predict IHC staining patterns directly from H&E-stained slides. These techniques aim to identify subtle morphological patterns that encode molecular information beyond human perception, offering a promising avenue for augmenting pathological analysis.

The advent of whole slide imaging, which converts histopathology glass slides into high-resolution digital images using specialized scanners, has enabled the application of DL techniques for automated histopathology analysis. DL algorithms, which can autonomously learn and extract meaningful features from high-dimensional data with minimal human intervention, are well-suited for computational pathology tasks[6].

Recent studies have demonstrated the potential of DL models to predict IHC staining patterns directly from H&E-stained whole slide images (WSIs) without requiring actual IHC staining. For instance, Su et al. developed the H&E Molecular neural network (HEMnet), which used P53 IHC stains as molecular labels to train a cancer classifier on clinical histopathological images[7]. HEMnet achieved high accuracy in identifying colorectal cancer regions in WSIs. Similarly, Frascarelli et al. described a convolutional neural network (CNN)-based model capable of predicting P53 mutational status and its spatial distribution in breast cancer tissue[8]. Shamai et al. trained a CNN model on breast cancer WSIs paired with PD-L1 IHC stains to accurately predict PD-L1 status[9]. Liu et al. developed a deep CNN model using WSIs of neuroendocrine tumors to effectively predict and quantify Ki-67 IHC status[10].

Existing studies primarily rely on conventional CNN-based DL algorithms. However, WSIs often contain billions of pixels, far exceeding the size of natural images, which poses significant challenges for traditional CNNs in terms of GPU memory and storage. Recently, transformers have emerged as a transformative DL technology, particularly for efficiently handling sequential and contextual tasks. Vision Transformers (ViTs), inspired by advancements in large language models like ChatGPT, are increasingly applied to image classification and have often outperformed traditional approaches[11].

LongViT, a novel ViT architecture leveraging LongNet, efficiently processes gigapixel-sized WSIs by handling extremely long sequences. It captures both long and short-range dependencies while simplifying image preprocessing by directly accepting high-resolution WSIs as input[12]. Moreover, LongViT has demonstrated strong performance on pathology tasks such as automated analysis of Gram-stained slides[13]. Prov-GigaPath, a recently introduced open-source foundational AI model, was pretrained on 1.3 billion pathology image tiles from 171,180 WSIs across 30,000 patients and 31 tissue types[14]. Using a pathology-specific adaptation of LongNet, this model has also demonstrated high accuracy in cancer classification and other digital pathology tasks[15].

Despite its capabilities, the Prov-GigaPath model has not yet been leveraged for large-scale real-world pretraining and context modeling to predict IHC stain patterns from H&E-stained WSIs. Additionally, many previous studies rely exclusively on H&E-stained slides throughout the analysis pipeline, without incorporating paired IHC slides[8]. Among those that do use both, most require image registration, assigning labels to H&E patches based on their corresponding regions in the registered IHC slide[7,10]. However, registration is often challenging and imprecise, particularly in paired datasets that lack precise pixel-wise alignment. Misalignment can introduce artifacts that may degrade model performance.

This study introduces HistoStainAlign, a novel training framework that leverages the Prov-GigaPath model to generate comprehensive slide embeddings from H&E WSIs while incorporating complementary information from paired but unregistered IHC WSIs. The paired nature of the datasets enables comparison of slide embeddings from H&E and IHC stains, even when tissue regions are not perfectly aligned. This approach captures the complementary information provided by both stain types while ensuring that only H&E-stained slides are used during model evaluation, maintaining clinical significance. This study develops, trains, and validates deep learning models based on Prov-GigaPath using a novel training framework to predict common IHC stain patterns from H&E-stained WSIs. The framework is evaluated on gastrointestinal and lung tissue WSIs with three widely used IHC stains (P53, PD-L1, and Ki-67) to assess its robustness and generalizability.

## MATERIALS AND METHODS

### Overview

HistoStainAlign is a deep learning framework designed to predict IHC staining patterns from H&E whole-slide images at inference time. The framework uses a foundational model, i.e., Prov-GigaPath[14], to extract features and generate embeddings from WSI image patches and aggregate these embeddings into a comprehensive slide-level representation. Furthermore, to enhance performance, HistoStainAlign integrates three novel key loss functions during training: 1) an inter-modality loss to align embeddings between H&E and IHC modalities, 2) an intra-modality loss to maintain consistency within the H&E modality, and 3) a class-based loss to optimize stain-specific classification. The effectiveness of HistoStainAlign as a training framework is demonstrated across three distinct and widely used IHC stains on clinically relevant downstream classification tasks.

**Datasets for IHC Biomarker Detection Tasks from H&E**

To evaluate the performance of HistoStainAlign, three labelled datasets were utilized in this study, each designed for a specific binary classification task: detecting P53 staining, PD-L1 staining, and Ki-67 staining from H&E WSIs. All slides were digitized at either 20x or 40x magnification (corresponding to 0.50 µm/pixel and 0.25 µm/pixel, respectively) using a Leica Aperio AT2 scanner (Leica Biosystems, Wetzlar, Germany). More details about these datasets are provided in Table 1. The use of these datasets for the study was approved by the Dartmouth-Health Institutional Review Board, which waived the requirement for informed consent as all data were de-identified.

| Task | Classification Scheme | Total WSIs | Total Patients | WSI Count per Class |
|---|---|---|---|---|
| P53 Detection | Wild-type vs. Positive Expression | 180 | 128 | Wild: 132  Positive: 48 |
| PD-L1 Detection | Low Expression vs. High Expression | 181 | 181 | Low: 149  High: 32 |
| Ki-67 Detection | Low vs. Intermediate/High Expression | 61 | 58 | Low: 41  Intermediate/High: 20 |

**Table 1:** Summary of datasets used in this study for binary classification tasks. For each IHC marker, the table lists the classification scheme, total number of available WSIs, total number of patients, and class distribution.

*Detection of P53 from H&E WSIs*

The P53 dataset consists of paired H&E-stained and IHC-stained FFPE biopsy samples collected from patients who underwent endoscopic distal esophagus and/or gastroesophagus junctional mucosal biopsies for detecting or surveillance of Barrett's esophagus at Dartmouth-Hitchcock Medical Center, a tertiary academic hospital in Lebanon, NH, between 2020 and 2023.

Board-certified anatomic pathologists categorized the WSIs into three staining patterns: wild-type staining, positive staining, and null staining. Due to the limited number of null staining patten samples, (n=13), this category was excluded from the study, resulting in a binary classification task distinguishing wild-type staining pattern from positive staining pattern samples. After data cleaning (removing 38 cases with no Barrett's H&E tissue and 10 without P53 staining slides), the final dataset included 180 WSIs from 128 patients.

*Detection of PD-L1 from H&E WSIs*

The PD-L1 dataset consisted of paired H&E-stained and IHC-stained FFPE tissue samples from patients who underwent lung tumor resection at Dartmouth-Hitchcock Medical Center between 2018 and 2023. PD-L1 expression score (Tumor Proportion Score or TPS) for each case was extracted from the corresponding pathology report and recorded as a percentage. According to Agilent Technologies, the TPS represents the percentage of viable tumor cells showing partial or complete membrane staining for PD-L1 relative to all viable tumor cells in the sample[16]. For this analysis, a binary classification scheme was adopted: samples with PD-L1 TPS score <50% were classified as low expression, while those with PD-L1 TPS score ≥50% were classified as high expression[16]. This approach was chosen to simplify classification, accommodate the small dataset, and ensure sufficient label representation. While some studies use a similar 50% cutoff[17–19], others have adopted alternative thresholds such as <1%, 1–49%, and >50%[20–22]. After data cleaning (83 slides with absence of lung cancer, 4 with mismatched H&E/IHC-stained slides, and 3 without TPS results were excluded), the dataset included 181 WSIs from distinct patients.

*Detection of Ki-67 from H&E WSIs*

The Ki-67 dataset consisted of paired H&E-stained and IHC-stained FFPE tissue samples collected from patients who underwent gastrointestinal biopsy or tumor resection at Dartmouth-Hitchcock Medical Center between 2022 and 2024. After data cleaning (removing 11 slides with missing or mismatched IHC slides), the dataset included 61 WSIs from 58 patients. The Ki-67 index, extracted from the corresponding pathology report and recorded as a percentage, was initially categorized into low expression (<3%), intermediate expression (3-20%), and high expression (>20%). The three-tier categorization corresponds to the grading system of gastrointestinal neuroendocrine tumors with Ki67 proliferation index (grade 1 with <3%, grade 2 with 3-20%, and grade 3 with >20%). Due to the limited number of high-expression cases, the intermediate and high-expression groups were combined. A binary classification scheme was then adopted: low presence (<3%) and intermediate/high presence (≥3%).

*Auxiliary Public Ki-67 Dataset for Pre-training*

Due to the limited labeled Ki-67 dataset available in this study, the ACROBAT[23] public dataset was used to pre-train models with the HistoStainAlign framework for this IHC stain using an unsupervised learning approach. The ACROBAT dataset includes 4,212 WSIs from 1,153 female primary breast cancer patients, each with one H&E stained WSI and up to four IHC-stained WSIs for routine diagnostic markers (ER, PGR, HER2, Ki-67). For this study, 843 H&E WSIs and their corresponding Ki-67 IHC-stained WSI were used to enhance the model's feature learning for Ki-67 detection. The pre-trained model was then evaluated on the labeled Ki-67 dataset from Dartmouth-Hitchcock Medical Center.

**Data Preprocessing**

The preprocessing pipeline involved extracting patches from both H&E and IHC WSIs, followed by generating tile embeddings for each patch.

*Patch Extraction*

The patch extraction process, illustrated in Figure 1, followed a consistent workflow for both H&E and IHC slides, with distinct filtering techniques tailored to each staining type. Binary tissue masks were generated from downsampled WSIs at 0.62x magnification and subsequently upscaled to 20x magnification. From these regions, 224×224-pixel patches were extracted, restricted to areas with sufficient tissue coverage (at least 20% stained material).

For H&E slides, an open-source digital pathology library[24], was used to generate the tissue masks. Background regions were filtered out to isolate tissue areas for patch extraction. For IHC slides, custom preprocessing steps were developed using the OpenCV library to enhance tissue quality and contrast. These steps included grayscale conversion, contrast enhancement, binary segmentation via Otsu thresholding to isolate tissue regions. Morphological operations (i.e., opening and closing) were applied to reduce noise and artifacts. The resulting binary mask was then upscaled to match a 20x magnification level for patch extraction. A sliding window approach was used to extract non-overlapping 224×224-pixel regions ensuring that each contains a minimum of 20% stained material.

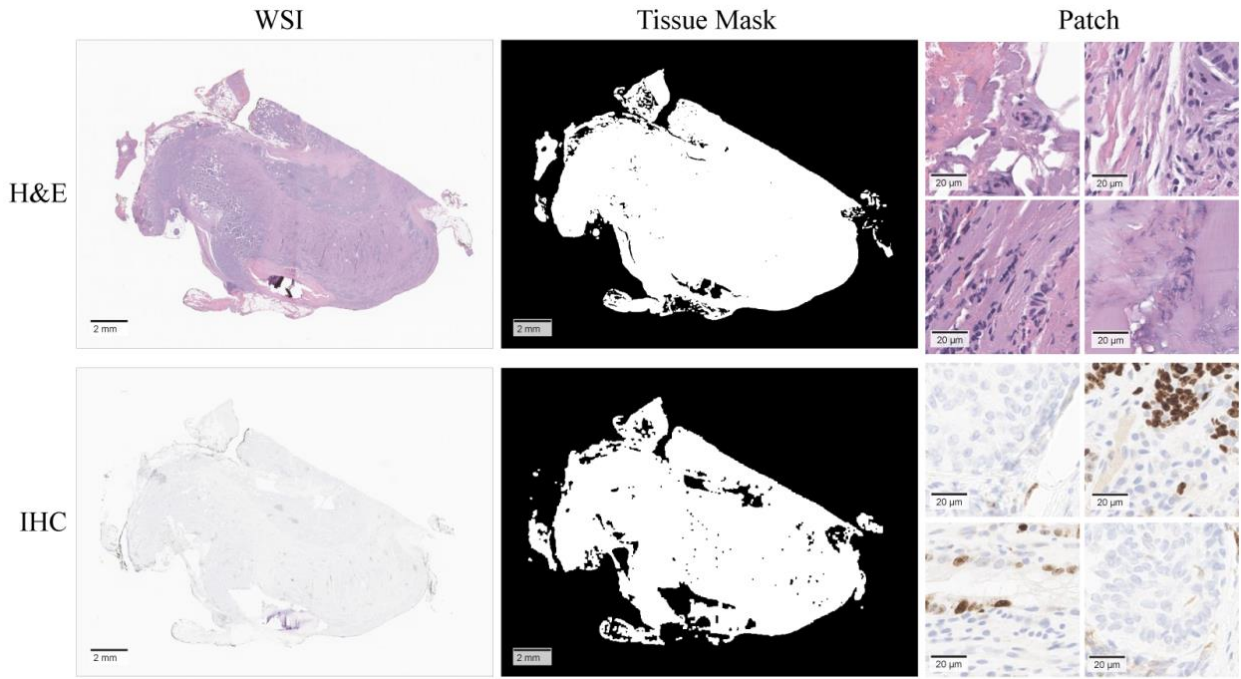

**Figure 1:** Example of tissue masks and corresponding patches extracted from H&E and IHC WSIs. Tissue masks were generated from downsampled WSIs and subsequently upscaled to 20x magnification. From these regions, 224×224-pixel patches were extracted, restricted to areas with sufficient tissue coverage (at least 20% stained material).

*Tile Embedding Generation*

After patches extraction, tile embeddings were generated using the Prov-GigaPath tile encoder, which has been pre-trained on 20x magnification H&E and IHC patches of size 224×224. The same encoder architecture and pre-trained weights were applied to H&E and IHC patches to maintain consistency across modalities. For each patch, the corresponding tile embedding and extraction coordinates were stored for downstream analysis.

**HistoStainAlign Framework**

HistoStainAlign is a deep learning framework that predicts IHC staining patterns from H&E WSIs by leveraging paired H&E and IHC data for contrastive learning during training. To generate slide-level representations, HistoStainAlign first divides each WSI into sequences of patches, with each patch represented by embeddings from the Prov-GigaPath ViT-based tile encoder. These patch embeddings are then aggregated using the Prov-GigaPath LongNet-based slide encoder to produce slide-level representations. Since this encoder was not pre-trained on paired H&E and IHC WSIs, it was hypothesized that fine-tuning it with paired slides would improve performance on the defined classification downstream tasks. HistoStainAlign's contrastive learning approach aligns H&E and IHC embeddings in a shared latent space rather than relying on precise pixel-wise registration. The framework is illustrated in Figure 2.

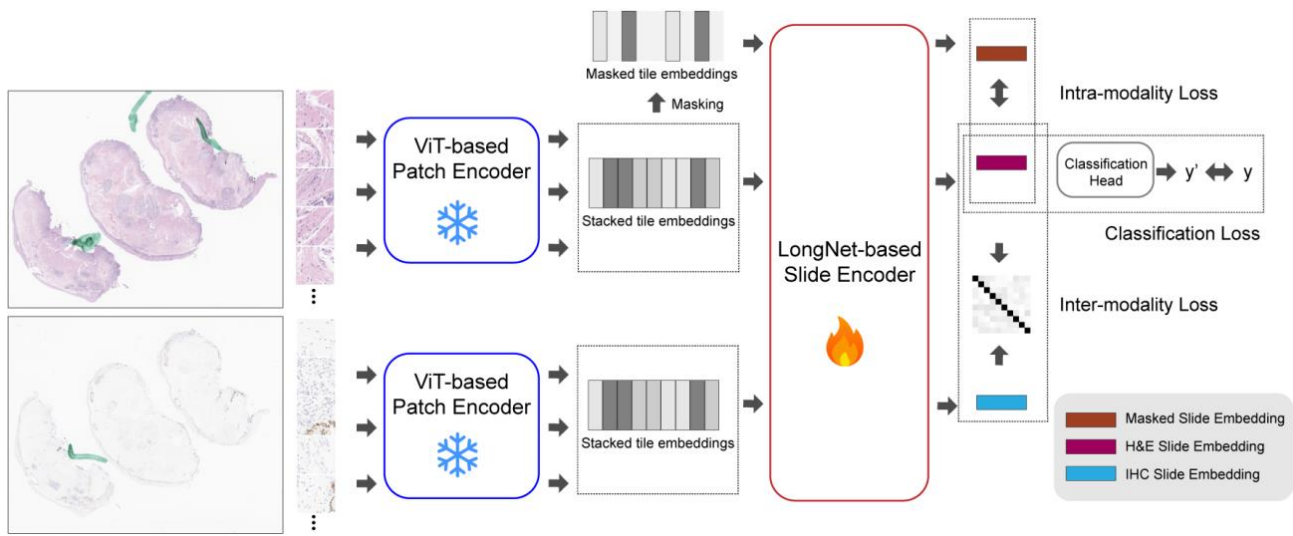

**Figure 2:** Architecture of the HistoStainAlign model. During fine-tuning, the model optimizes three types of loss: inter-modality, intra-modality, and classification loss. The ViT-based Patch Encoder (corresponding to the GigaPath tile encoder) processes H&E and IHC patches to generate tile-level embeddings, which are then aggregated by the LongNet-based Slide Encoder to produce slide-level representations. A contrastive loss is applied to the H&E and IHC slide embeddings to encourage alignment across modalities. Additionally, specific tiles are masked in the H&E patches to compute the intra-modality loss, promoting local consistency. The classification loss supports the downstream classification task.

*Loss Functions*

HistoStainAlign training is guided by three loss functions: inter-modality loss, intra-modality loss, and class-based loss. The inter- and intra-modality losses are self-supervised, while the class-based loss is supervised.

The inter-modality loss ($L_{inter}$) aligns the embedding spaces of paired H&E and IHC WSIs. A symmetric contrastive loss, adapted from TANGLE[25] (a transcriptomics-guided multimodal contrastive learning framework), it minimizes the distance between paired embeddings while maintaining separation between unpaired embeddings. Formally, given a batch of $N$ paired H&E and IHC WSIs, let $q_i$ represent the embedding of the $i$-th H&E WSI and $k_i^+$ the corresponding embedding of its paired IHC WSI. The inter-modality loss is defined as:

$$L_{inter} = \frac{1}{2N} \sum_{i=1}^{N} [CE(\frac{q_i \cdot K^+}{\tau}, y_i) + CE(\frac{k_i^+ \cdot Q}{\tau}, y_i)]$$

where $CE$ is the cross-entropy function.

The intra-modality loss ($L_{intra}$) ensures consistency within a single modality by leveraging masked embeddings. During training, 50% of patches in each sequence are randomly masked. Both the masked ($X_m$) and unmasked ($X$) sequences are passed through the slide encoder $f_{encoder}$, generating embeddings $Z = f_{encoder}(X)$ and $Z_m = f_{encoder}(X_m)$, respectively. The intra-modality loss minimizes the mean squared error (MSE) between these embeddings:

$$L_{intra} = L_{MSE}(Z, Z_m)$$

The class-based loss ($L_{class}$) optimizes predictions for downstream tasks. Using only H&E embeddings, the classification head generates predictions for the target label $y_i$. Given an embedding $e_i$ derived from the H&E WSI $q_i$ using the model $M$, the class-based loss is defined as:

$$L_{class} = CE(e_i, y_i)$$

The total loss function guiding HistoStainAlign training combines these three components:

$$L_{total} = L_{inter} + L_{intra} + L_{class}$$

By leveraging both self-supervised and supervised objectives, HistoStainAlign aligns H&E and IHC WSI embeddings, enforces intra-modality consistency, and optimizes performance classification tasks.

*HistoStainAlign-SSL: A Self-Supervised Variant*

HistoStainAlign-SSL is self-supervised variant of the standard HistoStainAlign framework, designed to learn robust slide representation without relying on labeled data. This variant excludes the classification head and optimizes only the inter-modality and intra-modality losses, aligning paired H&E and IHC embeddings while ensuring consistency through masked embedding reconstruction. By omitting the supervised class-based loss, HistoStainAlign-SSL is well-suited for scenarios with limited or no labels, such as the public Ki-67 dataset used in this study, as well as for pre-training models to enhance slide-level feature learning before downstream classification tasks.

**Evaluation**

To benchmark HistoStainAlign, its performance was compared against two transformer-based architectures that have demonstrated strong results in histopathology tasks: CONtrastive learning from Captions for Histopathology (CONCH) and UNI. The CONCH[26] model is pre-trained on over one million image-caption pairs, enabling it to learn rich multimodal representations. For classification, a zero-shot approach was employed in which textual descriptions of the target classes were paired with tile embeddings extracted from WSIs. These embeddings were then aggregated via top-k pooling to generate slide-level predictions. This strategy allows the CONCH to classify WSIs without task-specific fine-tuning and has been shown to outperform several state-of-the-art methods in histopathology. The UNI[27] model is pre-trained on more than 200 million pathology images across both H&E and IHC modalities. In the baseline comparison, the UNI tile encoder was used to generate embeddings for individual patches extracted from H&E WSIs. These tile embeddings were then aggregated using a Multiple Instance Learning (MIL) framework, which constructs slide-level representations that serve as inputs for a classifier. By leveraging large-scale pre-training, UNI facilitates robust slide-level predictions in histopathology.

To assess the performance of the HistoStainAlign framework, a 5-fold nested cross-validation approach was used. For each fold, the dataset was split into 60% training, 20% validation, and 20% testing, ensuring that all slides from the same patient were assigned to the same fold to prevent data leakage. After training, a logistic regression model was fitted to the training set of embeddings. Performance metrics, including area under the receiver operating characteristic curve (AUC), F1 score, precision, and recall, were calculated. Bootstrapping was used to compute 95% confidence intervals, providing robust statistical estimates. Results were aggregated across all test folds from the five cross-validation iterations, and confusion matrices were generated to visually assess classification performance.

To evaluate the contribution of the intra- and inter-modality loss functions, a model was trained using only the class-based loss. This baseline framework, named GigaPath-Finetuned, was assessed using the same cross-validation procedure to ensure consistency and enable a fair comparison. Cosine similarity was also calculated between paired H&E and IHC WSIs, as well as between unrelated H&E and IHC WSIs for both GigaPath-Finetuned and HistoStainAlign to analyze the impact of the loss functions on embedding alignment. The performance of the HistoStainAlign and GigaPath-Finetuned frameworks was evaluated across multiple metrics. GigaPath-Base, which uses the base Prov-GigaPath slide encoder without any further training, was also included as an additional baseline for comparison.

## RESULTS

### Model Performance

HistoStainAlign, which incorporates paired IHC and H&E slides during training but relies solely on H&E slides for evaluation, was compared against the GigaPath-Finetuned, which optimizes only the class-based loss. Results for models trained using the baseline methods, GigaPath-Base, GigaPath-Finetuned, and HistoStainAlign are shown in Table 2, while the corresponding confusion matrices for GigaPath-Finetuned and HistoStainAlign are presented in Figure 3.

| Task | Model | AUC | F1 | Precision | Recall |
|---|---|---|---|---|---|
| P53 Detection | CONCH-Base | 0.432 [0.325 - 0.527] | 0.680 [0.600 - 0.749] | 0.589 [0.503 - 0.667] | 0.585 [0.503 - 0.660] |
| | UNI-Base | 0.579 [0.483 - 0.668] | 0.638 [0.562 - 0.707] | 0.526 [0.451 - 0.611] | 0.523 [0.457 - 0.594] |
| | GigaPath-Base | 0.708 [0.618 - 0.792] | 0.695 [0.629 - 0.763] | 0.611 [0.536 - 0.692] | 0.613 [0.536 - 0.695] |
| | GigaPath-Finetuned | **0.750 [0.667–0.827]** | 0.711 [0.637–0.780] | 0.631 [0.549–0.715] | 0.631 [0.552–0.718] |
| | **HistoStainAlign** | 0.726 [0.635–0.812] | **0.735 [0.670–0.799]** | **0.662 [0.582–0.741]** | **0.666 [0.587–0.745]** |
| PD-L1 Detection | CONCH-Base | 0.497 [0.383 - 0.607] | 0.682 [0.619 - 0.750] | 0.588 [0.531 - 0.653] | 0.647 [0.556 - 0.742] |
| | UNI-Base | 0.620 [0.498 - 0.729] | 0.798 [0.731 - 0.865] | 0.666 [0.554 - 0.785] | 0.607 [0.528 - 0.688] |
| | GigaPath-Base | 0.711 [0.604 - 0.822] | 0.799 [0.739 - 0.859] | 0.656 [0.575 - 0.745] | 0.667 [0.578 - 0.767] |
| | GigaPath-Finetuned | 0.722 [0.623–0.822] | 0.806 [0.747–0.867] | 0.667 [0.581–0.755] | 0.683 [0.587–0.785] |
| | **HistoStainAlign** | **0.780 [0.679–0.877]** | **0.830 [0.772–0.886]** | **0.707 [0.618–0.799]** | **0.712 [0.621–0.808]** |
| Ki-67 Detection | CONCH-Base | 0.573 [0.422 - 0.732] | 0.582 [0.444 - 0.702] | 0.539 [0.415 - 0.661] | 0.542 [0.403 - 0.670] |
| | UNI-Base | 0.544 [0.395 - 0.693] | 0.568 [0.427 - 0.693] | 0.504 [0.368 - 0.637] | 0.504 [0.379 - 0.630] |
| | GigaPath-Base | 0.590 [0.424 - 0.757] | 0.672 [0.555 - 0.792] | 0.628 [0.499 - 0.766] | 0.628 [0.499 - 0.763] |
| | HistoStainAlign-SSL | 0.621 [0.455-0.790] | 0.705 [0.591-0.821] | 0.665 [0.540-0.793] | 0.665 [0.541-0.797] |
| | GigaPath-Finetuned | **0.672 [0.512 - 0.830]** | 0.695 [0.565 - 0.812] | 0.659 [0.515 - 0.791] | 0.640 [0.514 - 0.772] |
| | **HistoStainAlign** | 0.663 [0.503 - 0.823] | **0.723 [0.607 - 0.836]** | **0.686 [0.564 - 0.806]** | **0.690 [0.568 - 0.819]** |

**Table 2:** Performance comparison of multiple frameworks across three classification tasks: P53, PD-L1, and Ki-67 detection. Evaluated models include baseline methods (CONCH-Base and UNI-Base), GigaPath-Base (pre-trained Prov-GigaPath slide encoder), GigaPath-Finetuned (fine-tuned Prov-GigaPath encoder), and HistoStainAlign (fine-tuned Prov-GigaPath encoder using inter-

modality, intra-modality, and class alignment losses). For the Ki-67 detection task, HistoStainAlign-SSL, a self-supervised variant trained with additional Ki-67-stained slides from the ACROBAT dataset, is also included. Metrics reported are AUC, weighted F1-score, precision, and recall, with 95% confidence intervals provided in brackets. **Bold** values indicate the best performance, and <u>underlined</u> values indicate the second-best performance for each metric.

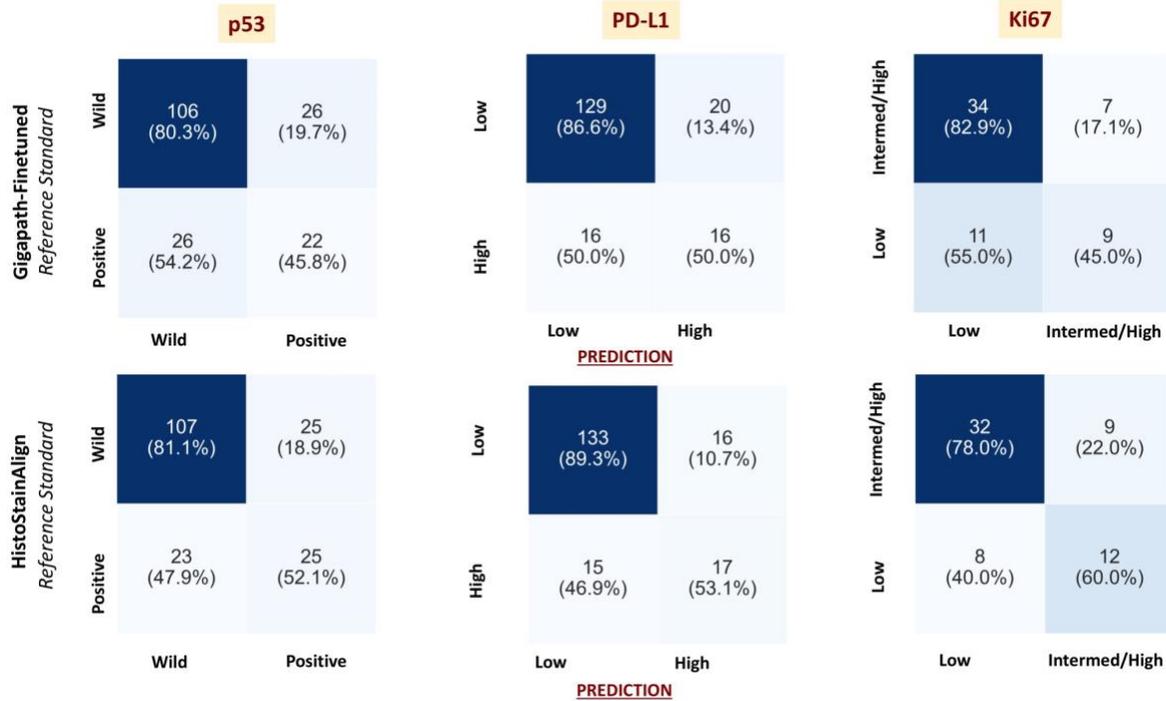

**Figure 3:** Confusion matrices for the GigaPath-Finetuned (top row) and HistoStainAlign (bottom row) models, presented from left to right for the P53, PD-L1, and Ki-67 datasets. The results are aggregated across the test set for all folds, providing a detailed overview of the model's performance for each dataset.

**Statistical Comparison of Performance of Models**

A Wilcoxon signed rank test for the prediction scores of true positive samples showed that HistoStainAlign significantly outperformed GigaPath-Finetuned in predicting PD-L1 status ($p<0.05$) but showed non-significant improvements for P53 ($p=0.32$) and Ki-67 ($p=0.34$). The self-supervised HistoStainAlign-SSL, pre-trained on the ACROBAT dataset, also showed non-significant improvement ($p=0.17$).

**Cosine Similarity Analysis**

The GigaPath-Finetuned model trains the slide encoder using only the class-based loss, whereas HistoStainAlign incorporates both inter- and intra-modality losses alongside the class loss. By leveraging symmetric contrastive loss, HistoStainAlign learns to distinguish similar and dissimilar pairs, aligning features between paired H&E and IHC WSIs during training.
To better understand the learned slide embeddings, cosine similarity between embeddings of WSIs were analyzed. For each model, paired H&E and IHC WSIs from the test set were used to compute cosine similarity, averaged across all folds. For each H&E WSI, cosine similarity was also computed with the remaining unpaired IHC WSIs in the test set introducing several random pairs. The results are summarized in Table 3. Embeddings generated by a model trained with HistoStainAlign show higher cosine similarity between paired slides compared to unpaired slides, with this difference being significantly more pronounced than in GigaPath-Finetuned across the P53, PD-L1, and Ki-67 datasets. These findings suggest that HistoStainAlign effectively incorporates paired H&E and IHC information into its learned embeddings, enhancing alignment across staining modalities.

| Task | Model | Paired Cosine Similarity (A) | Shuffled Cosine Similarity (B) | Difference (A - B) |
|---|---|---|---|---|
| P53 Detection | GigaPath-Finetuned | 0.901 [0.893-0.908] | 0.884 [0.876 – 0.893] | 0.017 [0.013-0.020] |
| | HistoStainAlign | 0.755 [0.747-0.763] | 0.602 [0.593 - 0.610] | 0.153 [0.142-0.164] |
| PD-L1 Detection | GigaPath-Finetuned | 0.667 [0.612-0.723] | 0.634 [0.578-0.690] | 0.033 [0.027-0.039] |
| | HistoStainAlign | 0.747 [0.742-0.753] | 0.596 [0.590-0.603] | 0.151 [0.145-0.157] |
| Ki-67 Detection | GigaPath-Finetuned | 0.774 [0.759-0.789] | 0.691 [0.672-0.710] | 0.083 [0.072-0.093] |
| | HistoStainAlign | 0.753 [0.742-0.765] | 0.599 [0.586-0.613] | 0.154 [0.140-0.168] |

**Table 3:** Cross-modality alignment analysis comparing GigaPath-Finetuned and HistoStainAlign models on P53, PD-L1, and Ki-67 tasks. Paired Cosine Similarity: Computed between embeddings of correctly matched H&E and IHC whole-slide images (WSIs), reported as mean ± 95% confidence interval (CI) across all paired slides. Shuffled Cosine Similarity: Computed as the average similarity between each H&E slide and all non-matching IHC slides within the test set (mean ± 95% CI). Difference: For each slide, the difference between paired and shuffled similarities (paired – shuffled) quantifies cross-modality alignment strength. Scores are aggregated across slides (mean ± 95% CI). A Wilcoxon signed-rank test showed that HistoStainAlign achieves significantly stronger alignment than GigaPath-Finetuned by comparing per-slide difference scores.

**DISCUSSION**

This study proposes a novel approach for predicting IHC staining patterns from H&E WSIs in gastrointestinal and lung tissue. Unlike existing methods that primarily focus on non-gastrointestinal cancers, such as breast and prostate, or rely solely on H&E WSIs, HistoStainAlign integrates paired IHC and H&E information without requiring patch-level annotations or tissue registration. By leveraging the state-of-the-art Prov-GigaPath model and training with slide-level labels, HistoStainAlign simplifies the training process while effectively utilizing all available data.

The results demonstrate that the HistoStainAlign framework effectively integrates morphological and molecular features across modalities. Cosine similarity analyses confirmed stronger alignment between paired H&E and IHC embeddings compared to shuffled pairs, supporting the effectiveness of the symmetric contrastive loss. Additionally, the model consistently outperformed baseline and fine-tuned GigaPath variants across all classification tasks, showing notable improvements in both alignment quality and predictive performance.

Furthermore, the HistoStainAlign-SSL framework, trained on the public ACROBAT breast cancer dataset, demonstrated improved performance over the GigaPath slide encoder in the Ki-67 detection task. Notably, this improvement was achieved despite being evaluated on gastrointestinal disease samples, even though the ACROBAT dataset is derived from breast cancer cases. This cross-domain generalization highlights the framework's ability to learn transferable features, suggesting its potential for broader applicability across different tissue types and disease contexts.

The baseline models used for comparison, CONCH and UNI, did not perform as well as those trained with the GigaPath model. This discrepancy is likely due to CONCH and UNI being primarily pretrained on H&E WSIs with minimal inclusion of IHC WSIs. Since accurately predicting IHC staining patterns from H&E images requires capturing relevant IHC-associated features, the additional IHC data in the GigaPath pretraining likely contributed to its superior performance. Additionally, the GigaPath model incorporates a dedicated slide encoder that aggregates tile embeddings into robust slide-level representations, whereas the UNI-Base model relies on a simpler MIL approach. The enhanced slide-level pretraining enabled by the GigaPath slide encoder further contributed to its improved performance. Given these advantages, the HistoStainAlign framework was built upon the GigaPath tile and slide encoders.

Although the performance improvements over the GigaPath-Finetuned model were not statistically significant for P53 and Ki-67 detection, the results remain promising. They highlight the potential of HistoStainAlign to further enhance performance even when building upon a strong baseline architecture. Notably, for PD-L1 detection, the framework achieved statistically significant gain in identifying clinically relevant positive samples ($p < 0.05$). The lack of significance in the other tasks may be attributed to the relatively small dataset sizes, yet the consistent performance trends across all targets suggest that HistoStainAlign successfully leverages paired H&E and IHC data to improve predictive alignment. These findings reinforce the strength of the underlying foundational model's architecture while also demonstrating a principled way to build upon it for modality-aware classification.

Integrating this framework into clinical practice could help mitigate the high cost and labor/time demands associated with traditional IHC staining. Currently, IHC staining increases the cost of biopsy by nearly 50% based on Medicare reimbursement rates and requires hours of technical processing, along with additional time for pathological interpretation. An automated deep learning system like the one proposed in this study, capable of rapidly and accurately predicting various IHC stains, has the potential to serve as a cost-effective diagnostic aid for pathologists.

One limitation of this study is the lack of large, paired datasets of H&E and IHC WSIs, which constrains the full potential of multimodal learning. Future work will focus on collecting larger paired datasets spanning diverse staining types and tissue regions. Such datasets would enable self-supervised pretraining to better align H&E and IHC modalities, followed by fine-tuning for specific diagnostic tasks. Pretraining on larger paired datasets could improve the robustness and generalizability of the framework across different clinical settings. Additionally, while the performance metrics achieved by HistoStainAlign are promising, they leave room for improvement. These limitations are likely attributable to the small size and limited diversity of the current datasets. Expanding the dataset to include greater variability in staining patterns and tissue types may significantly enhance the model's performance. Also in this study, the Tumor Proportion Score (TPS) was used for grading PD-L1 for binary classification with grading of 50% as the cut off for low- and high-level expression. Currently, TPS equal or more than 1% is a standard cut-off for positivity used in clinical trials and treatment guidelines, while 50% is often considered as higher PD-L1 expression. This high-level expression is associated with a better response to immunotherapy[28]. In the future, different grading classifications—including 1%—can be explored using a larger number of WSIs.

Despite these limitations, this study presents a simplified and robust framework for multimodal integration in computational pathology. By eliminating the need for tissue registration and pathologist-provided patch-level annotations, it offers a scalable approach that maximizes the utility of available data. Extending this methodology to additional IHC marker stains and cancer types could further broaden its clinical applicability. With continued refinement, the proposed framework has the potential to advance digital pathology by improving diagnostic accuracy and optimizing the use of resource-intensive staining procedures.

## ETHICS STATEMENT AND PATIENT CONSENT

## ACKNOWLEDGMENTS

The authors would like to thank Laura J. Gordon for her invaluable assistance with data collection.

**DATA AVAILABILITY STATEMENT**

The model and code for HistoStainAlign will be made publicly available upon publication. The datasets used and/or analyzed during the current study consist of paired H&E and IHC whole-slide images and are available from the corresponding author upon reasonable request, subject to institutional review board agreement.

**Declaration of Generative AI and AI-assisted Technologies in the Writing Process**

During the preparation of this work, the authors used ChatGPT (OpenAI, San Francisco, CA) solely for proofreading to improve the clarity and readability of the manuscript. The authors reviewed any suggested edits, modified the text as needed, and take full responsibility for the content of the publication.

**Author Contributions**

A.D., N.T., XL and S.H. performed study concept and design; A.D., N.T., B.R., X.L., and S.H. performed development of methodology and writing, review and revision of the paper. K.S., W.M., P.C., K.C. provided acquisition and interpretation of data. All authors reviewed and approved the final paper.